\documentclass[a4paper,11pt]{article}
\usepackage{jinstpub}
\usepackage{xcolor}
\usepackage{graphicx}
\usepackage{makecell}
\usepackage[colorlinks=true]{hyperref}
\usepackage{amsmath}
\usepackage{multirow}
\usepackage{float}
\usepackage{lineno}
\usepackage{subcaption}
\usepackage{pifont}
\usepackage{caption}
\usepackage{array}
\title{Data reduction strategy in the PandaX-4T experiment}
\author[a]{Yubo Zhou}
\author[a,b,1]{Xun Chen,\note{Corresponding author.}}

\affiliation[a]{INPAC and School of Physics and Astronomy, Shanghai Jiao Tong University, MOE Key Lab for Particle Physics, Astrophysics and Cosmology, Shanghai Key Laboratory for Particle Physics and Cosmology,\\ Shanghai 200240, China}
\affiliation[b]{Shanghai Jiao Tong University Sichuan Research Institute,\\ Chengdu 610213, China}
\affiliation[ ]{(PandaX-4T Collaboration)}

\emailAdd{chenxun@sjtu.edu.cn}
\abstract{ The PandaX-4T experiment is designed for multiple purposes,
  including searches for solar neutrinos, weakly interacting massive
  particles, and rare double beta decays of xenon isotopes. The
  experiment produces a huge amount of raw data that needs to be
  stored for related physical analyses in a wide energy range. With
  the upgrading of the PandaX-4T experiment, the doubled sampling rate
  resulted in a larger data size, which challenges both the cost and
  the data processing speed.  To address this issue, we propose a data
  reduction strategy by removing the noise tail of large signals and
  downsampling the remaining parts of them. This strategy reduces the
  requirement for storage by 65\% while increasing data processing
  speed. The influences on physical analyses on different topics at
  different energy regions are negligible.  }
\begin{document}

\maketitle

\section{Introduction}
The PandaX-4T experiment~\cite{PandaX:2018wtu}, located in the China
Jinping Underground Laboratory (CJPL)~\cite{Li:2014rca}, is a
multi-purpose program for the study of particle physics and
astrophysics.  Its physics goals include the search of weakly
interacting massive particles (WIMPs)~\cite{PandaX-4T:2021bab}, solar
neutrinos~\cite{PandaX:2022aac}, and the rare double beta decay of
xenon isotopes~\cite{PandaX:2022kwg}.  The experiment utilizes the
technology of dual-phase xenon time projection chamber
(TPC)~\cite{RevModPhys.82.2053} to catch the signals generated by
these events.

The PandaX-4T detector is introduced in detail in
Ref.~\cite{PandaX-4T:2021bab}. The main part of the TPC is nearly
cylindrical and assembled with 24 highly reflective
polytetrafluoroethylene (PTFE) panels. The TPC primarily contains
liquid xenon, with a thin layer of gas phase at the top. The drift
field in the liquid xenon is determined by the cathode grid on the
bottom of the TPC and the gate mesh beneath the top surface of the
liquid xenon. The electric field to extract electrons from liquid to
gaseous phase is created using the gate mesh and the anode mesh in the
gaseous xenon. Two arrays of 3-inch photo-multipliers (PMTs) are
located at the top and bottom of the TPC to measure the photons
produced inside the TPC. The PMTs amplify the photoelectrons, which
are produced by incoming photons via the photoelectric effect on the
photocathode, by a factor of about $10^6$.

When particles scatter with xenon atoms in the sensitive volume, they
can produce both atomic excitation and electron ionization. The
de-excitation of xenon atoms and the recombination of some of the
ionized electrons produce prompt scintillation or the so-called
``$S1$'' signal when detected by the PMT arrays. The remaining ionized
electrons drift along the drift field to the top surface of liquid
xenon and are extracted into the gas region to produce the
proportional electroluminescence or ``$S2$'' signal if detected. Given
the nearly constant drifting velocity of electrons inside the liquid
xenon, the time difference between the $S1$ and $S2$ signals can be
used to determine the depth of the scattering event inside the
sensitive volume~\cite{RevModPhys.82.2053}.

In the Pandax-4T experiment, the $S1$ and $S2$ signals are
reconstructed from the collected data. For each PMT, the output signal
is digitized at a sampling rate of 250~MHz. When the digitized signal
exceeds a predefined threshold, a continuous segment of the samples,
including those over a pre-defined threshold and a fixed number of
samples before and after them, is recorded~\cite{Yang:2021hnn}. The
raw data is the collection of recorded segments, together with the
time and channel information.

\begin{figure}[htb]
  \centering
  \includegraphics[width= 0.8 \textwidth]{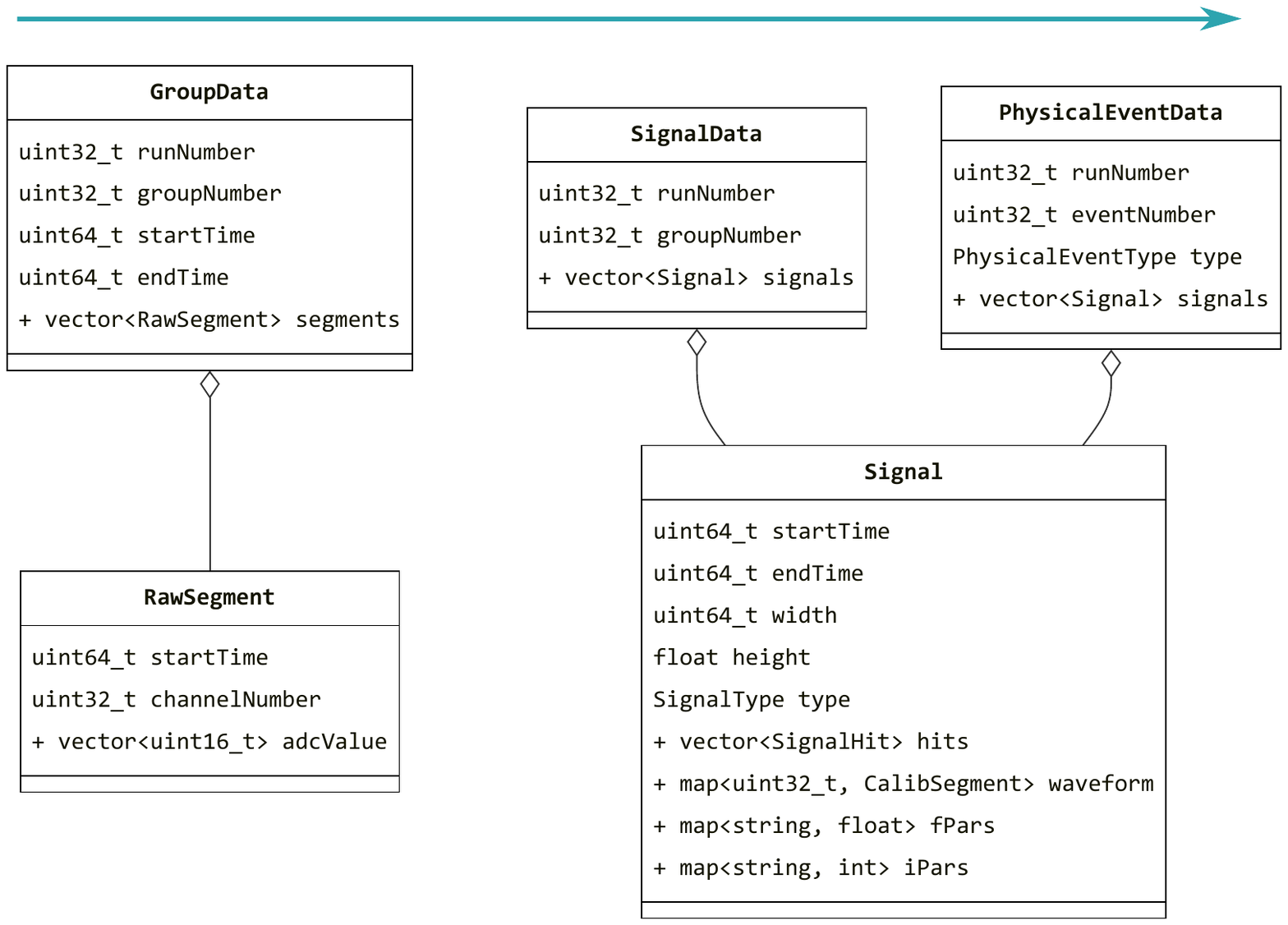}
  \caption{The custom structure of data in PandaX-4T. The
    \texttt{RawSegment} class contains the samples (\texttt{adcValue}), the
    start time of the first sample and the channel number. The
    \texttt{GroupData} class consists of the run number, the group
    number, time of the first sample and last sample inside it and a
    list of \texttt{RawSegment}s. During data processing, a
    \texttt{GroupData} is converted to a \texttt{SignalData}
    structure, which contains all the reconstructed signals in the
    group. The \texttt{PhysicalEventData} structures are created by
    breaking up the boundary of the groups and combining related
    signals.}
  \label{fig:bs3_data}
\end{figure}

For efficient input/output (I/O) and data compression, the raw data in
the PandaX-4T experiment are stored using the Bamboo Shoot
3~\cite{PandaX:bamboo-shoot3} library, which supports custom
formatting and compression of ordinary data structures. As a result,
the size of a data file in the PandaX-4T experiment is approximately
half that of the actual data size contained within it. The fundamental
unit of the custom data structure is called \texttt{RawSegment}, which
accurately represents the raw samples, channel number and start time.
Many \texttt{RawSegments} collected within a short period of time are
grouped together, creating a higher-level structure known as
\texttt{GroupData}, as depicted in Figure~\ref{fig:bs3_data}. Each
\texttt{GroupData} is characterized by a start time and an end time,
which define the group boundary. The start time is defined as the
start time of the first \texttt{RawSegment}, while the end time
corresponds to the end time of the last finished
\texttt{RawSegment}. To ensure separation between adjacent
\texttt{GroupData}s, a minimum gap of 1$\mu$s is required.  A single
data file may contain thousands of \texttt{GroupData} objects.

For each data file, the data processing is performed by looping over
the \texttt{GroupData} structure. A baseline is determined for each
segment and then subtracted. The resulted waveforms are calibrated
using light-emitting diodes (LEDs). Peaks in the waveform are
identified as hits, and adjacent hits are clustered together to form
signals. The top waveform of a signal is defined as the sum of the
waveforms of all its hits in the top array. Similarly, the bottom
waveform is the sum of the waveforms of hits in the bottom array. The
total waveform is defined as the combined sum of the top and bottom
waveforms.  The signals are tagged as either $S1$, $S2$ or other types
based on their properties, such as height, width, and top-bottom
asymmetry of charge partition. The identified signals are grouped as
\texttt{SignalData}, with the same boundary as its parent
\texttt{GroupData}, as shown in Fig.~\ref{fig:bs3_data}.

In order to reconstruct physical events, it is necessary to eliminate
the boundary between adjacent \texttt{SignalData} blocks, as a single
physical event may contain signals that are separated across them. To
address this, two methods have been
developed~\cite{PandaX4T:signal_model_paper}. The first method
involves iterating through the signals in time order. During this
process, the maximum $S1$ signal is recorded prior to encountering any
$S2$ signals, and the iteration concludes when the maximum drift time
to the maximum $S1$ is reached. Once found, this signal, along with
the preceding and subsequent signals within a drift time window of
1~ms, is combined to form a physical event. This approach has been
utilized in all of the currently published PandaX-4T
results~\cite{PandaX-4T:2021bab, PandaX:2022kwg, PandaX:2022aac,
  pandax2023limits, PhysRevLett.129.161803,
  PhysRevLett.129.161804}. The second method is developed to improve
the reconstruction efficiency of physical events in low energy region
during the ongoing combined analysis with both the commissioning run
(Run0, 2020-2021) and the first science run (Run1, 2021-2022) data.
This method begins by identifying an $S2$ signal that does not have
any $S2$ signals surpassing one-third of its total charge within a
2~ms window prior to it.

Following the process of identifying an S2 signal, a physical event 
constructed by selecting all the signals within a 1~ms window before
and after the identified $S2$ signal. The signals belonging to a
reconstructed physical event are stored within the data structure
called \texttt{PhysicalEventData}, and an event number is assigned to
uniquely identify this event in its containing data file.

The data containing the physical event is then converted to ROOT format
for further analyses. The waveforms of the signals are not kept in the
ROOT file due to the huge space consumption.

Throughout Run0 and Run1 of the PandaX-4T experiment, high-quality raw
data, including both background data and calibration data spanning
roughly 280 days, occupies approximately 750~TB. It can be
extrapolated that a year of data collection would lead to roughly 1~PB
of raw data accumulation.  Additionally in the coming data taking, the
implementation of a new electronics system upgrade will result in a
higher sampling rate of 500 MHz~\cite{He:2021sbc}.  The resulting
doubling of data during future operations of PandaX-4T represents a
significant burden on the limited storage capacity available. This
challenge has motivated the PandaX-4T team to actively seek solutions
to optimize the data storage and processing capabilities of the
experiment.

This article is organized as follows: We
discuss the data reduction strategy and its performance in
Sec.~\ref{sec:data_reduction}. The effects of data reduction on the
data analysis results at different energy scales are discussed in
Sec.~\ref{sec:physical_results}. The update of the data processing
chain is presented in Sec.~\ref{sec:chain}. At last, a brief summary
is given in Sec.~\ref{sec:summary}.

\section{Data reduction strategy}
\label{sec:data_reduction}

\subsection{Data information}
\begin{figure}[hbt]
  \centering
  \includegraphics[width= 0.6 \textwidth]{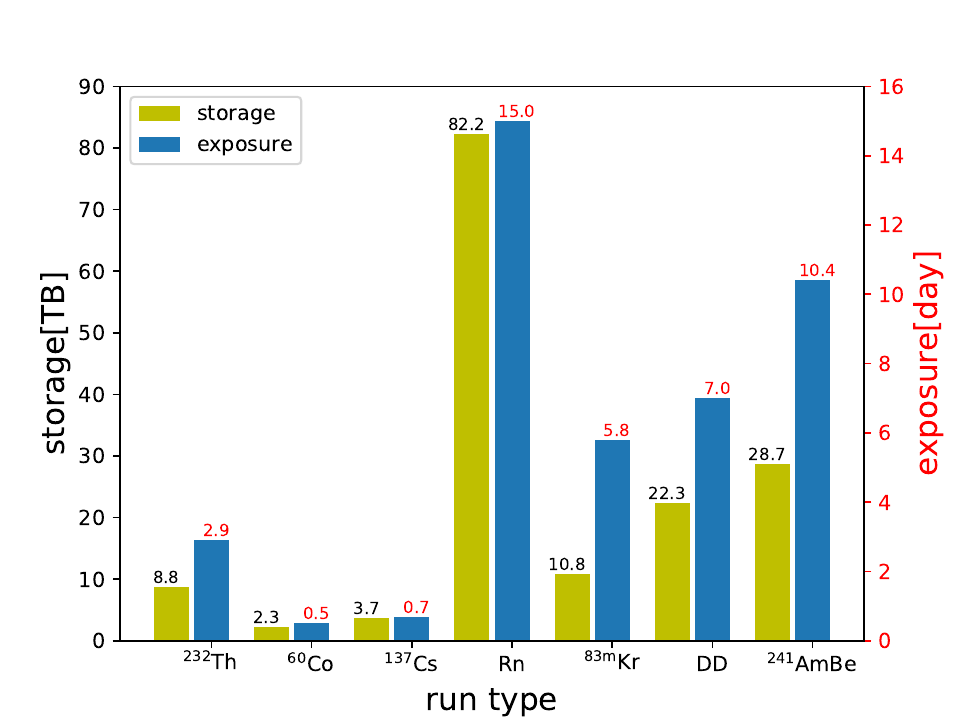}
  \caption{The storage occupation and accumulated run time of
    different types of calibration data. The dark matter search data,
    with 86.1\% of the total exposure time, occupies 67.8\% of total
    storage, and its corresponding information is not shown in this
    plot.}
  \label{fig:data_occupation}
\end{figure}

The development of a data reduction strategy starts from the study of
existing data. The total storage occupation of different types of raw
data and corresponding run time are plotted in
Fig~\ref{fig:data_occupation}. Then we calculated the average
occupation rates of data collected in the Run0 and Run1, by dividing
stored data size by data taking duration recorded in the database. The
information is summarized in Table.~\ref{tab:data_rates}. Higher rate
of the $^{232}$Th calibration run in Run1 is a result from different
positions of the source, i.e., the middle loop in Run1 and the
deuterium-deuterium (DD) tunnel in Run0~\cite{Zhou:2019tevpa}. The
high event rate in $^{83\rm{m}}$Kr run of Run1 is mainly caused by the
Rn contamination introduced from the injection chamber and higher
injection flow. The increasing of background rate in Run1 can be
attributed to the elevated level of Rn background. Among all the
calibration runs, Rn runs exhibit the highest average data production
rate. This can be explained by the presence of a larger number of high
energy events in Rn runs, resulting in more large $S2$ signals with
longer persistence times and an increased number of samples.

\begin{table}[hbt]
  \centering
  \small
  \begin{tabular}{c|ccccccc|c}
    \hline
    \multirow{2}{*} {Data type} & \multicolumn{7}{c|}{calibration} &\multirow{2}{*} {background}\\
     & $^{232}$Th & $^{60}$Co & $^{137}$Cs & Rn & $^{83\rm{m}}$Kr & DD & $^{241}$AmBe & \\\hline
    Run0 & 47.3 & 59.2 & 65.1 & 36.5 & 16.1 & 36.0 & 28.9 & 11.2  \\
    Run1  & 64.8 & 58.0 & 52.8 & 91.3 & 51.9 & 41.8 & 40.3 & 18.1  \\
    \hline
  \end{tabular}
  \caption{Average occupation rates of different type of science data in PandaX-4T, in MB/s.}
  \label{tab:data_rates}
\end{table}

The LUX collaboration found that electron emission occurring within a
time range of $\mathcal{O}(10)$~ms after $S2$ pulses is a notable
background in the low-energy region~\cite{LUX:2020vbj}.  To mitigate
the impact of those $S2$ afterglow, a cut is implemented that excludes
events occurring within a 22~ms window after a large $S2$ signal with
a magnitude larger than 10,000 PE~\cite{PandaX-4T:2021bab}.

%A large $S2$ pulse is defined as an
%$S2$ signal larger than 10,000~PE.

\begin{figure}[hbt]
   \centering
   \includegraphics[width= 0.6 \textwidth]{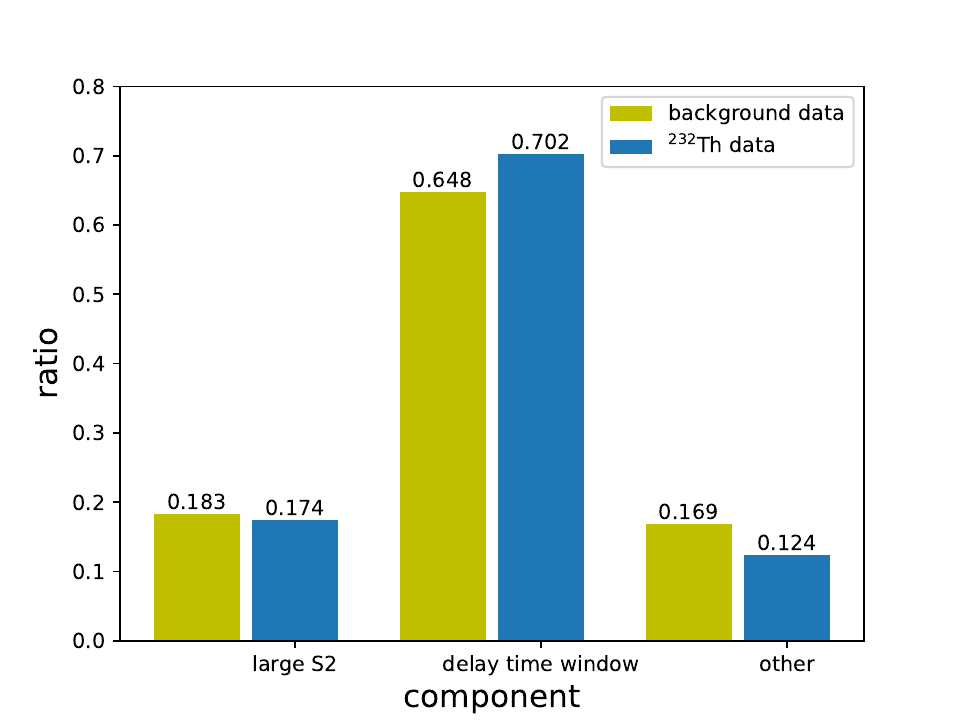}
   \caption{The storage occupation ratio of different components of data.}
   \label{fig:sig_type_occupation}
\end{figure}

The signals discarded in the afterglow veto cut could be a major
factor in the storage of collected data.  To test this hypothesis, we
selected 100 files from the dark matter search data and 100 files from
the $^{232}$Th calibration data randomly in the Run0. We then counted
the signals in the 10~ms window after a large $S2$ pulse, which we
refer to as the delay time window. As shown in
Fig.~\ref{fig:sig_type_occupation}, data in the delay time window
contributes to about 2/3 of the storage space and large $S2$ pulses
themselves account for nearly 1/5.
 
\subsection{The reduction procedures}
The investigation of data occupation suggests that significant storage
space reduction is possible by removing the unused portions after
large $S2$ pulses and reducing the sampling rate of these pulses. The
procedure of removing these portions is referred to as ``tail cut'',
while the process of reducing the sampling rate is known as
``downsampling'' in the subsequent text.

To minimize the impact on the calculation of some quantities for event
selection, such as the extra charge except main signals, an algorithm
has been developed to determine the start and end time points for the
tail cut procedure. The detailed description of this algorithm is
provided below.

\begin{figure}[H]
  \centering
  \begin{subfigure}[b]{0.4\textwidth}
    \includegraphics[width=\textwidth]{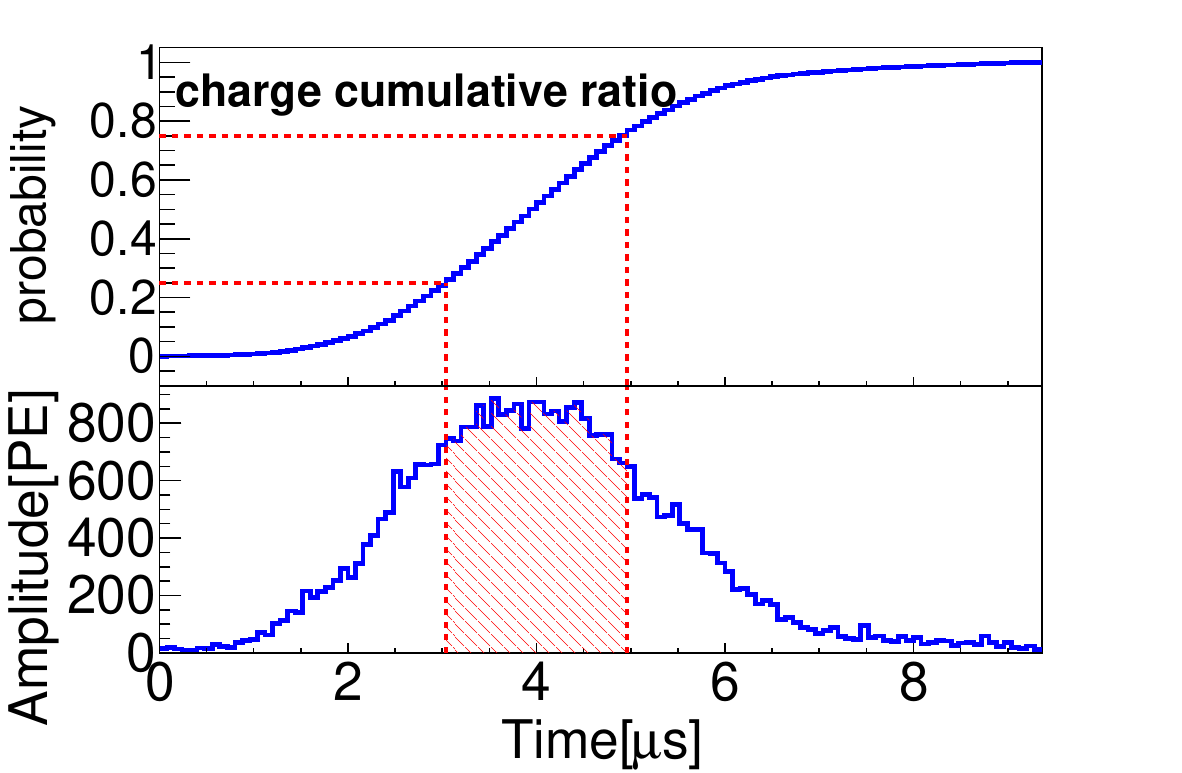}
    \caption{Definition of ${\eta}_{ccw}$}
    \label{fig:ccw_def}
  \end{subfigure}
  \begin{subfigure}[b]{0.4\textwidth}
    \includegraphics[width=\textwidth]{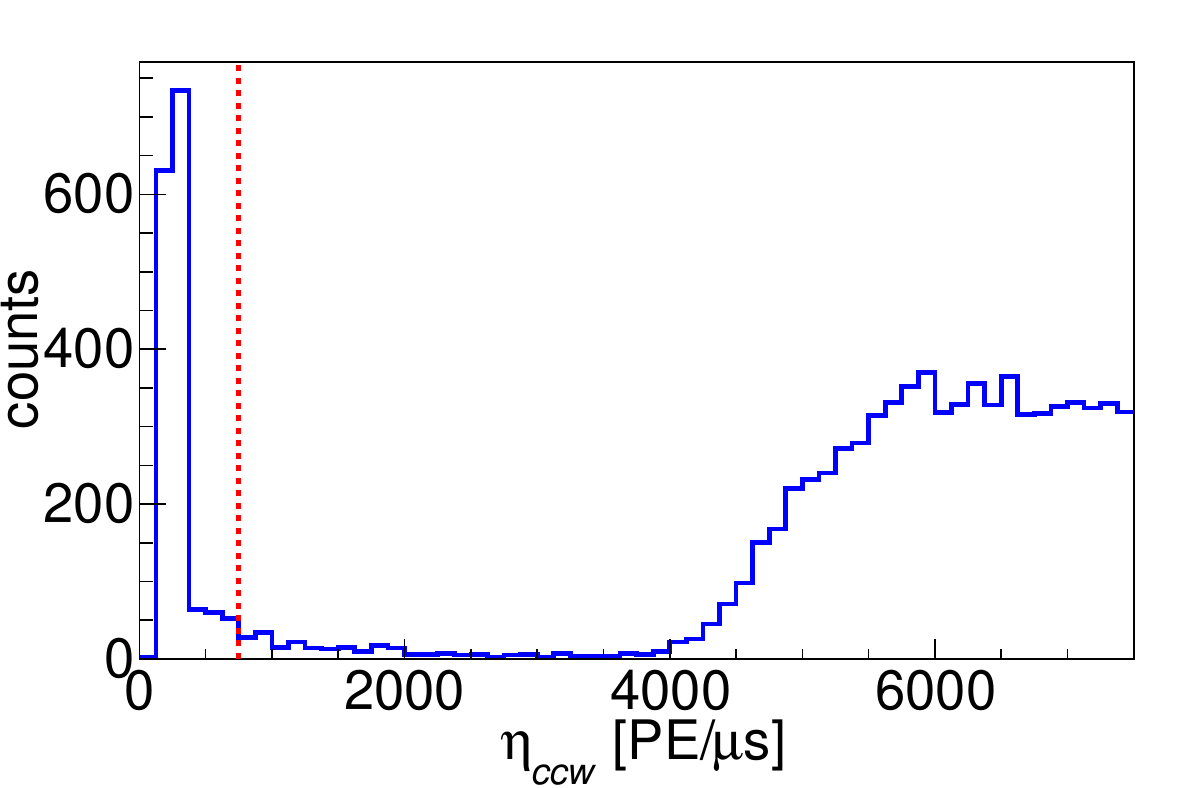}
    \caption{Distribution of ${\eta}_{ccw}$}
    \label{fig:ccw_dis}
  \end{subfigure}
  \caption{The definition(\subref{fig:ccw_def}) and
    distribution(\subref{fig:ccw_dis}) of
    ${\eta}_{ccw}$. \subref{fig:ccw_def}) The red shadow area occupies
    central 50~\% charge. ${\eta}_{ccw}$ is defined as the ratio
    between area and corresponding width.  \subref{fig:ccw_dis}) The
    red dashed line indicates the cut used to distinguish
    mis-identified $S2$s (with low ${\eta}_{ccw}$) from ordinary
    ones.}
  \label{fig:ccwratio}
\end{figure}

First, the original data are processed by the general data processing
chain to construct signals. The time of the constructed $S2$ pulses
will be recorded. These $S2$s will then be investigated further to
remove the mis-identified pulses using the distribution of the
${\eta}_{ccw}$ variable, which is the charge-to-width ratio of central
50\% waveform. Fig.~\ref{fig:ccwratio} shows the definition and
distribution of this variable. The mis-tagged backgrounds have
relatively low ${\eta}_{ccw}$ in comparison with good $S2$s.

Remaining $S2$ pulses will be clustered together if the time
difference between the start times of adjacent $S2$ pulses is within
1~ms. Each cluster of $S2$ pulses will be treated as a whole. Only
clusters with at least one $S2$ pulse greater than 20,000~PE will be
considered in the following steps.

\begin{figure}[!htbp]
  \centering
  \includegraphics[width=0.95\textwidth]{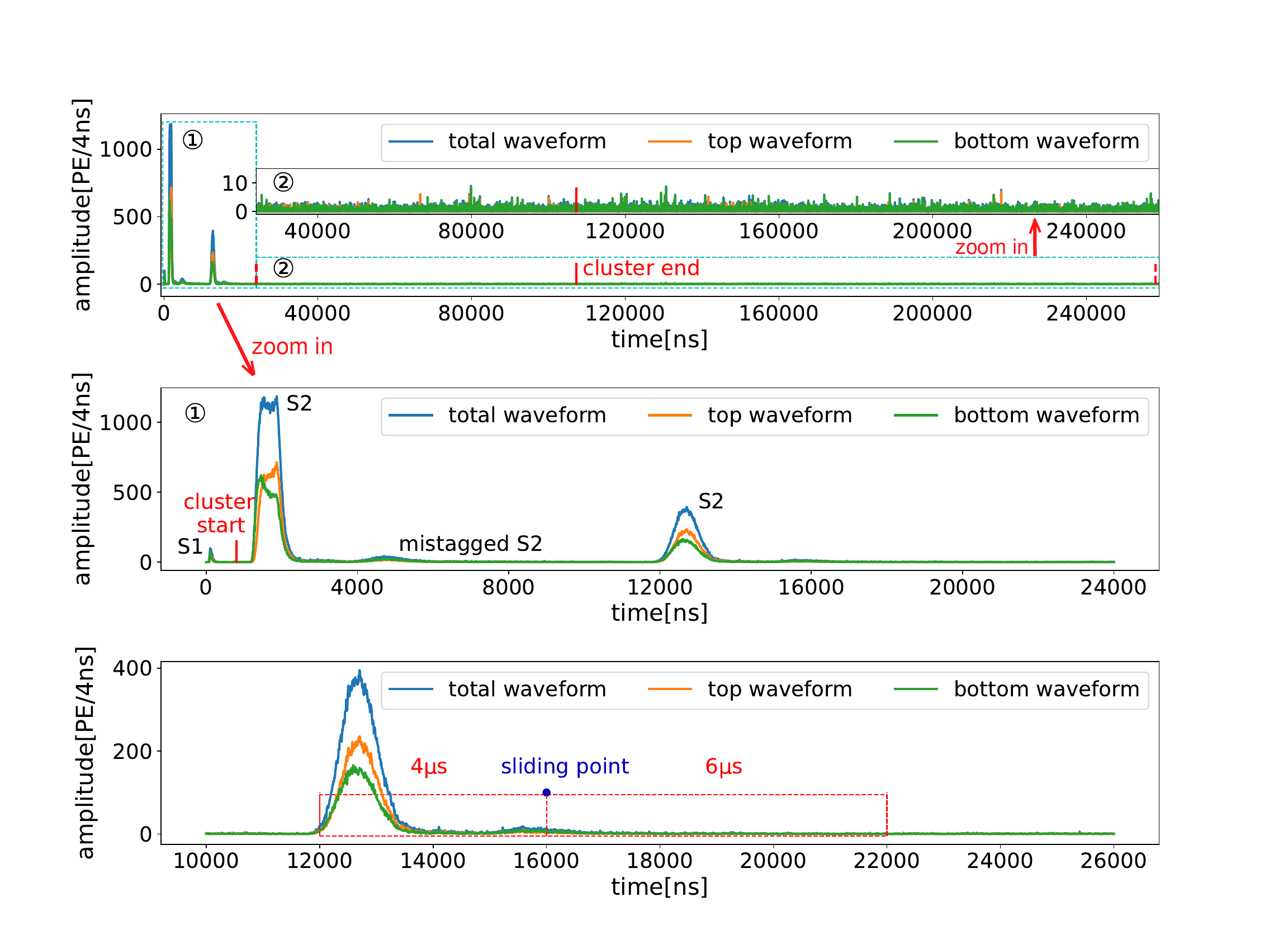}
  \caption{An example to demonstrate the tail cut algorithm. A
    complete waveform with tail cut region is shown in the panel
    above. Several major $S2$ pulses are identified in region
    \ding{172}, as shown in the middle panel. Region \ding{173} is the
    tail part to be removed. The red solid bars indicate the boundary
    of the $S2$ cluster. The red dashed bars mark the start and end of
    the tail cut part.  A sliding point is depicted in the bottom
    panel. The quantities of $\bar{Q}_{pre4}$ and $\bar{Q}_{post6}$ are
    computed within the red outlined regions before and after the
    sliding point, respectively. The point moves backward along the
    time axis in steps of 2~$\mu$s. }
  \label{fig:p_downsample_area}
\end{figure}

An example demonstrating the determination of the tail cut range is
shown in Figure~\ref{fig:p_downsample_area}. For each cluster, a
reverse sliding point is set on the waveform, starting from the end of
the last $S2$ pulse and moving backwards in steps of 2~$\mu$s.
The step size of 2~$\mu$s is selected to strike a balance between
processing speed and the rate of misidentification of S2s with
magnitudes around 1,000 PE. This choice yields comparable results to
using smaller step sizes, but with faster processing time. On the
other hand, larger step sizes significantly increase the rate of
misidentification, leading to the loss of valuable information.
At each step, the average charge per sample is calculated in the 4~$\mu$s
window before the sliding point ($\bar{Q}_{pre4}$) and the 6~$\mu$s
window after it ($\bar{Q}_{post6}$). The sliding stops when
$\bar{Q}_{pre4}>2,500$~PE/$\mu$s or
$\bar{Q}_{pre4}>5\bar{Q}_{post6}$. This indicates that a large signal
is before the sliding point. The start point of the tail cut is set at
8~$\mu$s after the final sliding point.

The end point of the tail cut is determined by iterating over the
reconstructed signals after the cluster of $S2$ pulses, rather than on
the waveform. Because larger $S2$s lead to more delay signals in a
longer time window, we adjust the iteration step size based on the
cluster's total charge to speed up the calculation, as shown in
Table ~\ref{tab:step_size}.  At each step, the ratio of the total
charge of the signals to the difference between the start time of the
first signal and the end time of the last signal in this step is
calculated. The iteration stops until the ratio is smaller than
0.5~PE/$\mu$s or an $S2$ pulse with a ${\eta}_{ccw}$ greater than the
threshold of 2,500~PE/$\mu$s is found in the step. The end time of the
last signal in the previous iteration is set as the end point of the
tail cut.

\begin{table}[hbt]
  \centering
  \small
  \begin{tabular}{c|c}
    step size& range of cluster total charge (PE)\\\hline
    10 & $(0, 5\times10^4)$ \\
    20 & $[5\times10^4, 10^5)$ \\
    40 & $[10^5, 1.5\times10^5)$ \\
    60 & $[1.5\times10^5, \infty)$ \\\hline
  \end{tabular}
  \caption{The step size of iteration to determine the end point of
    tail cut and the corresponding cluster total charge.}
  \label{tab:step_size}
\end{table}

Reducing of the sampling rate (downsampling) of the remaining $S2$
pulses in the clusters could suppress the storage space occupation
further. The updated sampling rate is 125~MHz. The downsampling is
performed on the data structure of \texttt{RawSegment}.  The raw
\texttt{adcValue} is split into small groups of size $n$, where $n$ is
the ratio between the original and target sampling rates.  For the
data collected during Run0 and Run1, the original sampling rate was
250 MHz, resulting in a value of $n=2$. However, with the anticipated
upgrade of PandaX-4T, the upcoming data will be acquired at a sampling
rate of 500 MHz, leading to an increased value of $n=4$.  The
downsampled \texttt{adcValue} is calculated by averaging the values in
each group. To reduce the round-off error caused by integer division, a 
value of $n/2$ is added to the summation of each group before averaging.

\subsection{Reduction performance}
To evaluate the performance of the data reduction strategy, 50 files
were selected for each type of data. The results are shown in
Table~\ref{tab:reduction_performance}.  Applying the tail cut strategy
alone could reduce storage occupation by more than 50\%. Using
downsampling in addition to the tail cut strategy could save an
additional 10\% of disk space.  Due to the higher fraction of high
energy deposition, calibration data contains a significant amount of
data in the tail part, resulting in increased storage
requirements. Therefore, after the implementation of the reduction
strategies, calibration data has been further reduced.

\begin{table}[hbt]
  \centering
  \small
  \begin{tabular}{c|ccccccc|c}
    \hline
    \multirow{2}{*} {Data type} & \multicolumn{7}{c|}{Calibration} &\multirow{2}{*} {Background}\\
     & $^{232}$Th & $^{60}$Co & $^{137}$Cs & Rn & $^{83\rm{m}}$Kr & DD & $^{241}$AmBe & \\\hline
    raw & 1008.0 & 1004.7 & 1009.4 & 1003.8 & 1003.6 & 1007.3 & 1009.0 & 1011.2  \\
    tail cut only  & 419.8 & 414.5 & 483.1 & 421.9 & 377.7 & 409.7 & 420.3 & 460.2  \\
    tail cut and downsampling & 308.4 & 297.2 & 338.5 & 323.9 & 297.1 & 305.5 & 309.9 & 364.0  \\
    \hline
  \end{tabular}
  \caption{Average data storage occupation (MB) of a single data file
    before and after data reduction.}
  \label{tab:reduction_performance}
\end{table}

\section{Effect on data analysis}
\label{sec:physical_results}
Data reduction permanently removes some data, which may affect total
exposure and event reconstruction. We discuss the impact of data
reduction on analysis in this section.

\subsection{Exposure}
The exposure is calculated by excluding regions determined by the
afterglow veto cut from the actual data taking time. For example, the
fixed veto cut reported in Ref.~\cite{PandaX-4T:2021bab} has resulted
in an exposure loss of 7.3\% for Run0 data. Recently, a new adaptive
veto cut is implemented, leading to a smaller exposure loss of
about 5.25\% when applied to data in Run0 and
Run1~\cite{PandaX4T:signal_model_paper}.

When the afterglow veto cut is combined with the data reduction, the
excluding region will be determined as the result from the ``union''
operation of the regions given by both algorithms.  Since the tail cut
algorithm removes data that would mostly be covered by the veto cut,
the loss of exposure introduced by the data reduction is small when
combined with the adaptive veto cut. The resulting exposure loss is
about 5.32\%.

\subsection{Large $S2$ pulses}

\begin{figure}[!thbp]
  \centering
  \includegraphics[width=\textwidth]{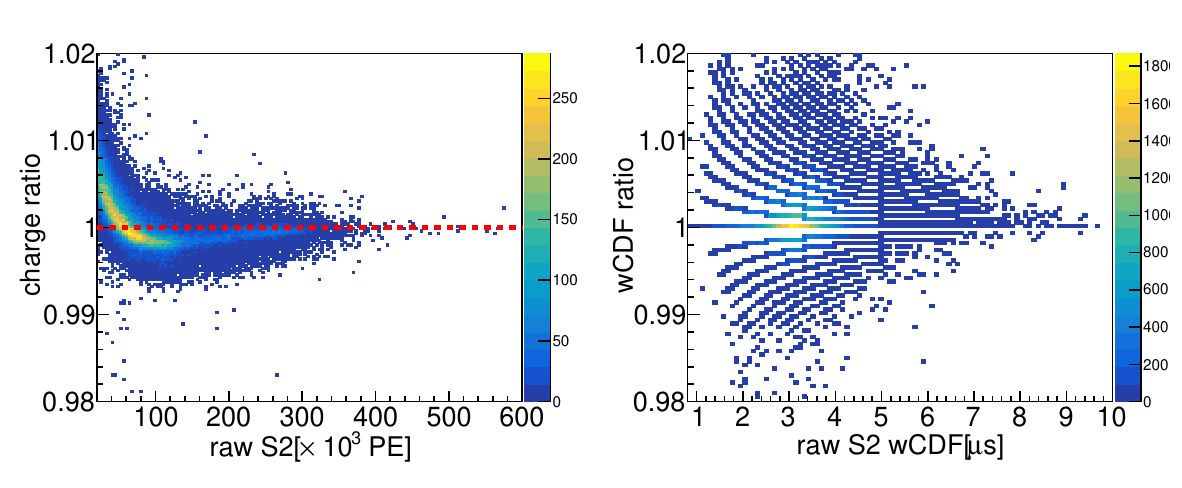}
  \caption{ Left: the ratio distribution of charge after reduction to
    that prior reduction for large S2 pulses. Right: the ratio
    distribution of wCDF after reduction to that prior reduction for
    large S2 pulses.  }
  \label{fig:p_S2_comparison}
\end{figure}

For large $S2$ pulses over 20,000 PE, the total charge and width of
the central 80\% charge (wCDF) are expected to change after data
reduction. wCDF is an important variable used to design data quality
cuts. To study these effects, we inspected the large $S2$ pulses in
single scattering (SS) events~\cite{PandaX:2022kwg} with all data
quality cuts applied in a $^{232}$Th calibration run. The results are
presented in Fig.~\ref{fig:p_S2_comparison}. The charge and wCDF
deviations are both smaller than 1\% for the vast majority of the
studied pulses (97.6\% and 95.3\%, respectively).

\begin{figure}[!thbp]
  \centering
  \includegraphics[width= 0.6\textwidth]{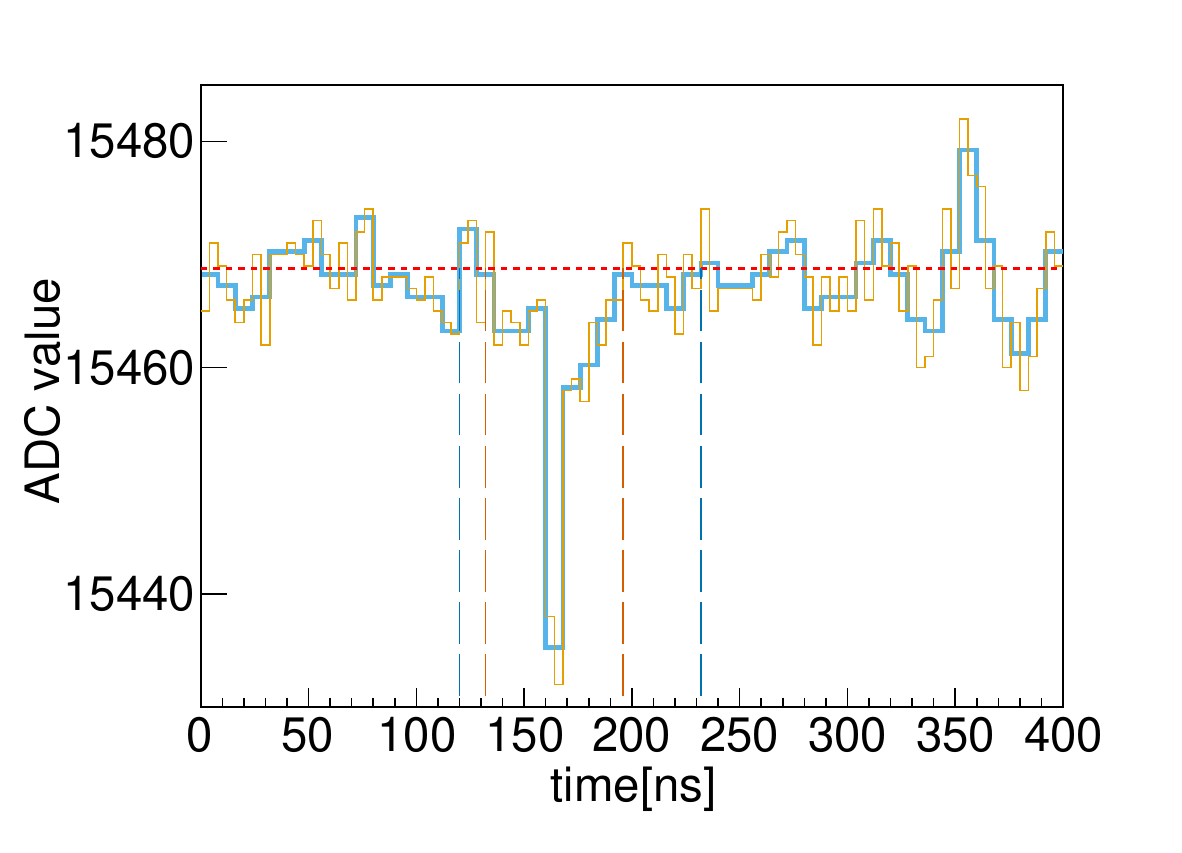}
  \caption{
  \label{fig:p_rawsegment}
  A ``fatter'' hit is recognized after downsampling from 250~MHz
  (orange dashed line) to 125~MHz (blue dashed line). The borders of the hit before
  and after the downsampling are marked. Baseline is calculated with the mean value of 
  front 20 samples before downsampling(red dashed line).}
\end{figure}

Notably, for raw $S2$ charge below approximately 50,000 PE, the
reduced charge typically increases, whereas for charges exceeding
50,000 PE, the reduced charge tends to decrease.  This phenomenon can
be explained by the combination of two opposing effects. The first
effect is attributed to downsampling. As illustrated in
Fig.~\ref{fig:p_rawsegment}, downsampling smooths waveform
fluctuations, leading to ``fatter'' small hits that contribute more
charge. On the other hand, the second effect arises from the tail cut,
which involves discarding noisy tails of large $S2$ pulses, resulting
in a reduction in charge. The downsampling effect has a greater impact
on smaller $S2$ pulses as they have a higher proportion of small hits,
leading to an increase in charge after reduction. Conversely, larger
$S2$ pulses tend to exhibit a decrease in charge due to these effects.

\subsection{Events in the mid-to-high energy region}
\begin{figure}[htb]
  \centering
  \begin{subfigure}[b]{0.8\textwidth}
    \includegraphics[width=\textwidth]{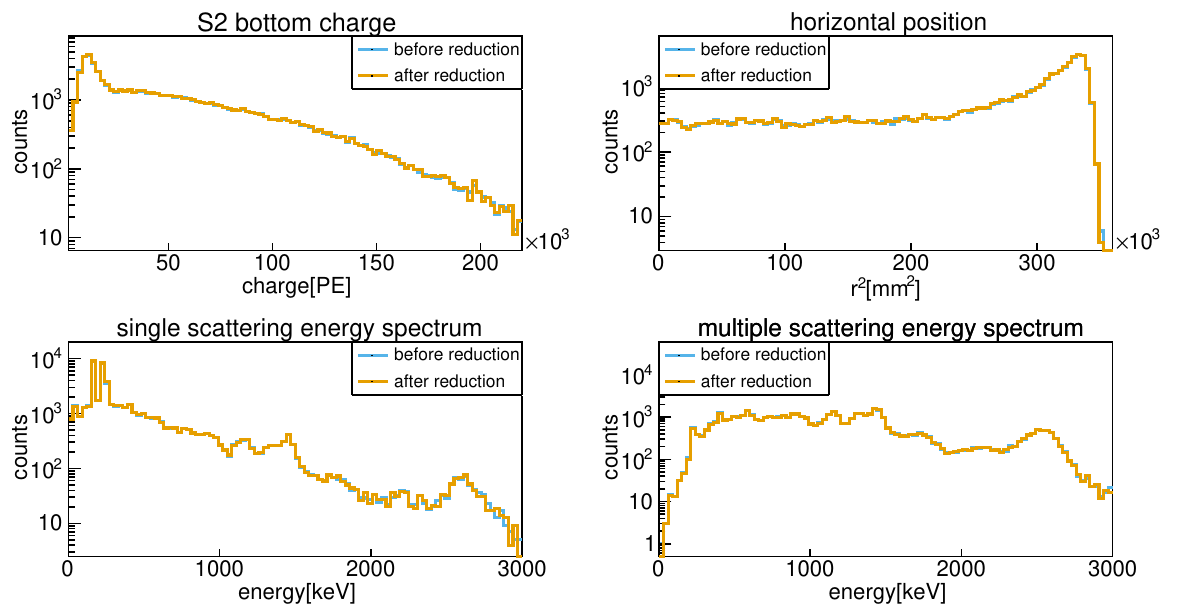}
    \caption{background run}
  \end{subfigure}
  \begin{subfigure}[b]{0.8\textwidth}
    \includegraphics[width=\textwidth]{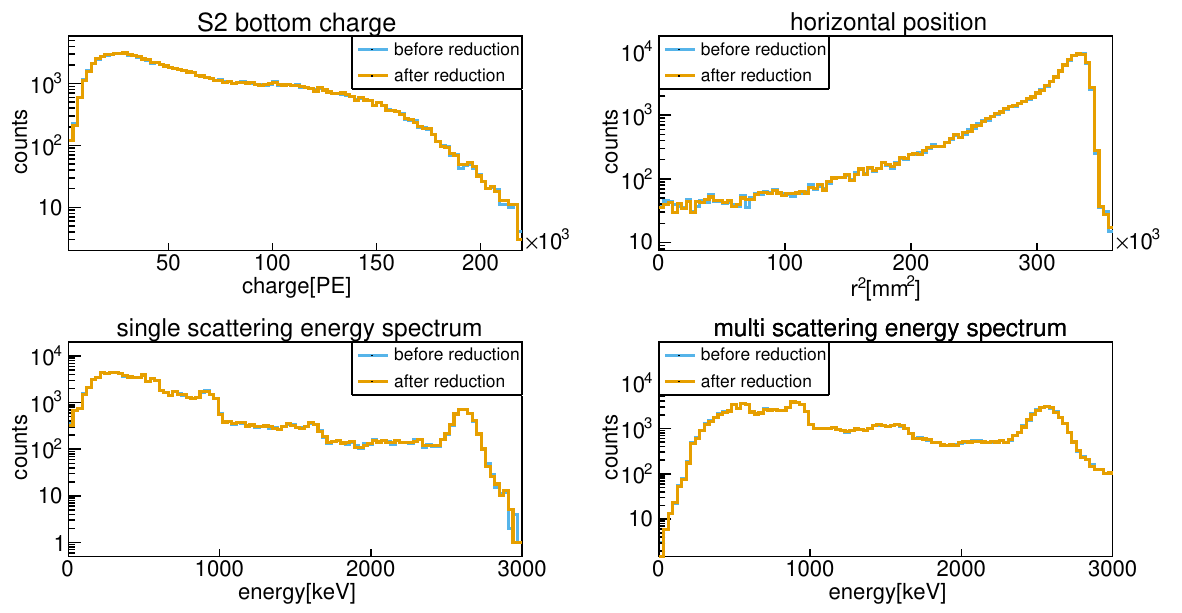}
    \caption{$^{232}$Th run}
  \end{subfigure}
  \caption{$S2$ bottom charge, reconstruction position and energy spectra
    before(blue) and after(orange) data reduction of two different types of
    runs.}
  \label{fig:p_spectrum_MHE}
\end{figure}

The mid-to-high energy region in PandaX-4T data analysis refers to the
energy range between 20~keV and 3000~keV, where many interesting
physical phenomena occur.  In addition to the $^{232}$Th run mentioned
previously, a background run was also used to test the effects of data
reduction on data analysis in the mid-to-high energy region.
Fig.~\ref{fig:p_spectrum_MHE} shows the distribution of $S2$ bottom
charge, horizontal position after desaturation~\cite{PandaX:2022kwg},
and energy spectra for SS events that passed the data quality cuts.
The energy spectra of multiple scattering (MS)
events~\cite{PandaX:2022kwg} are also shown.  We conclude that data
reduction has a negligible effect on the distributions.  In the
context of spectrum fitting technologies commonly employed in related
data analyses within this energy range, the minimal changes observed
in the spectra would have negligible effects on the fitting results.
Table~\ref{tab:number_event_data} shows the number of reconstructed SS
and MS events before and after data reduction.  The total number of
events changes slightly after the data reduction, mainly because some
adjacent events may be reconstructed as one after the tail cut. Some
of these events are recognized as MS, while others are filtered out
by data quality cuts.

\begin{table}[hbt]
  \centering
  \small
  \begin{tabular}{c|c|c|r}
    \hline
    Type/Counts & before reduction & after reduction & change \\ \hline
    background, SS & 7583  & 7558 & $-0.33\%$\\
    background, MS & 6820  & 6816 & $-0.06\%$\\
    $^{232}$Th, SS & 101209 & 100771 & $-0.43\%$\\
    $^{232}$Th, MS & 130479  & 130728 & $0.19\%$\\
    \hline
  \end{tabular}
  \caption{The number of reconstructed events in the mid-to-high
    energy region in selected runs, before and after the data
    reduction.}
  \label{tab:number_event_data}
\end{table}

\subsection{Events in the low energy region}
The low energy region is below 20~keV, where WIMPs and solar
$^8$B neutrinos are searched. We evaluated the effect of data
reduction on a Rn calibration run in Run1 firstly. After data
reduction, 2 of the 3,186 candidate events disappeared, and no new
candidate events appeared. Inspection of these two events showed that
the event reconstruction algorithm expanded their event window due to
the tail cut before them, introducing additional noise. As a result,
they were filtered out by the waveform cleanliness requirement.

To further evaluate the effect of data reduction, we selected only the
files containing candidate events after data selection and checked
other types of data. For the Rn data in Run0, only 1 of the
2,177 inspected candidate events was lost for the same reason as in
the Run1 Rn data. No new candidate events appeared. For the
neutron calibration data with $^{241}$AmBe and DD sources, the reduced data
files produced identical event lists to the original data files. Using
the same method, no candidates were lost among the 4,645 selected
candidates selected with basic quality cuts in Run0 for $^8$B
neutrinos search~\cite{PandaX:2022aac}. Additionally, there were no
newly emerged candidates. The analysis results in the low energy
region do not change after data reduction.

\section{Updates of the data processing framework}
\label{sec:chain}
Because data reduction has a negligible effect on data analysis in all
energy regions, it will be used in the upcoming PandaX-4T data
taking. This will require modifications to the previously used data
processing framework.

\subsection{Data structure}
The first thing to be updated is the underlying data storage
structure. The \texttt{RawSegment} structure assumes that all raw data
is sampled at 250 MHz. However, the new electronics sample data at
500 MHz, and downsampling in the data reduction will transform some
of the data to 125 MHz. To accommodate these changes, a new variable
called \texttt{sampleSize} is added to the structure to indicate the
size of samples in nanoseconds. The value of \texttt{sampleSize} is 4 for the
original Run0 and Run1 data, 2 for the upcoming raw data, and 8 for the
downsampled \texttt{RawSegment}.

\subsection{Multi-stage data processing}

Data reduction requires multiple processing stages. In the first
stage, the default data processing algorithm is applied to the raw
data to reconstruct signals. However, the adaptive afterglow veto cut
requires the signal densities in the noise tails after large $S2$
pulses, so these parameters must be calculated before data
reduction. Therefore, the second stage calculates these parameters and
stores them in the database, along with the meta information of runs
and files. The list of large $S2$ pulses with their time is also
generated in this stage. In the third stage, reduced data is produced
according to the information of large $S2$ pulses extracted in the
previous step. The reduced raw data is then processed again to
generate data for further
analysis. Table~\ref{tab:time_consuming_data} shows the average time
consumption of different data processing steps, calculated from 50
randomly selected background run data files.

\begin{table}[hbt]
  \centering
  \small
  \begin{tabular}{c|cccccc}
    \hline
    \makecell[c]{Process step/\\time(seconds)} & build signals & \makecell[c]{generate\\ parameters} & data reduction & \makecell[c]{build physical\\ events} & \makecell[c]{convert to\\ ROOT file} & \makecell[c]{event selection}\\ \hline
    before reduction & 67.63  & 31.40 & 21.76 & 49.80 & 187.19 & 14.72  \\
    after reduction & 29.14 & - & - & 17.66 & 49.82  & 13.46  \\
    \hline
  \end{tabular}
  \caption{Time consumption at each step of data processing before and after data reduction.}
  \label{tab:time_consuming_data}
\end{table}
The previous data processing chain took an average of 319.34 seconds
to convert a raw data file into data for physical analysis. The new
multi-stage data processing chain reduced the average required time to
230.87 seconds. More importantly, during the data reprocessing
campaign of reduced data, the average required time to process one
file was only 110.8 seconds.

\section{Summary}
\label{sec:summary}
We propose a data reduction strategy to save storage space in the
PandaX-4T experiment. The strategy was designed by analyzing the data
components and is capable of saving more than 60\% of storage
space. It has a negligible effect on data analysis in different energy
regions and can also greatly speed up data processing. 

The strategy is to be used in the following data taking operation
after the upgrade of the PandaX-4T experiment.  Additionally, 1/10 of
the raw data files will be preserved without any reduction applied to
serve as a reference for evaluating the impact of the reduction
strategy.  We hope that it will provide a useful reference to other
experiments with large amounts of data, and that it will help to
reduce the cost and complexity of data storage and processing for
these experiments.

\section*{Acknowledgments}
\label{sec:ack}
We would like to express our gratitude for the meticulous internal
review conducted by Prof. Qing Lin and Dr. Yi Tao. This project is
supported in part by the grants from National Science Foundation of
China(No. 12090060 and No. 12175139).

\bibliographystyle{unsrt}
\bibliography{reference/refs}

\end{document}